Ref:
Aljarboa, S. & Miah, S.J. (2020). An Integration of UTAUT and Task-Technology Fit Frameworks for Assessing the Acceptance of Clinical Decision Support Systems in the Context of a Developing Country, In the Proceedings of Fifth International Congress on Information and Communication Technology, ICICT 2020, London

# An Integration of UTAUT and Task-Technology Fit Frameworks for Assessing the Acceptance of Clinical Decision Support Systems in the Context of a Developing Country

Soliman Aljarboa[1] and Shah J. Miah[2]

[1] Victoria University Business School, VIC, Australia
[2] Newcastle Business School, University of Newcastle, NSW, Australia

**Abstract.** This paper is to create a basis of theoretical contribution for a new PhD thesis in the area of Clinical Decision Support Systems (CDSS) acceptance. Over the past three years we conducted the qualitative research into three distinctive phases to develop an extended Task-Technology Fit (TTF) Framework. These phases are for initiating requirement generation of the framework, discovering the factors of the framework through perspectives, and evaluating the new proposed framework. The new condition is related to developing country in which various sectors such as healthcare is mostly under attention. We conduct a new inspective for assisting decisions support technology and its usefulness in this sector to integrate with other frameworks for assisting the value, use and how can be better accepted in context of healthcare professionals.

## 1   Introduction

CDSS is one type of Health Information Systems (HIS) that is used in diagnoses, dispensing appropriate medications, making recommendations and providing relevant information that all contribute to medical practitioners' decision-making [1]. CDSS help medical practitioners to make their decisions and produce good advice based on up-to-date scientific proof [2]. CDSS is a system that needs more research for new knowledge generation to reconcile and increase the interaction between the physician and CDSS in order to assist and support the physician to use the system successfully.

It is necessary to investigate the determinants of CDSS acceptance for medical applications. According to Sambasivan et al. [3], developing countries face more difficulties than developed countries in implementing HIS. They also argued that improving the quality of healthcare performance can only be achieved by the acceptance of HIS by doctors. Several frameworks have been studied to determine the factors that affect acceptance of HIS. However, there is still a lack of research regarding the factors that affect physicians' acceptance of CDSS in developing countries [4].

Understanding and identifying the factors that influence the acceptance of information technology help designers and suppliers to reach a better understanding regarding the requirements of the end users. This leads to providing information systems which are more appropriate to the characteristics and conditions of the user and the work environment.

## 1.1 Unified Theory of Acceptance and Use of Technology (UTAUT)

The UTAUT model was established and examined by Venkatesh et al. [5], where they investigated and analyzed eight different models and theories in order to discover and identify the factors that influencing user acceptance of technology. These models included the following: the theory of reasoned action; the technology acceptance model; the motivational model; the theory of planned behaviour; a model combining the technology acceptance model and the theory of planned behaviour; the model of PC utilization; the innovation diffusion theory, and the social cognitive theory. This contributed to providing and providing a model capable of interpretation and gaining a greater understanding of user acceptance of technology. In addition to that, the factors most influencing user acceptance were also identified, and this led to many studies in several fields to use the UTAUT model.
The UTAUT model includes four variables: gender, age, experience and voluntariness of use. In addition UTAUT model comprises four major determinants which are:: performance expectancy, effort expectancy, social influence and facilitating conditions [5].

## 1.2 Task-Technology Fit (TTF)

TTF Model indicates that information technology has a positive and effective role on user performance in the event that the features and characteristics of the technology are appropriate and fit with the business mission [6]. The TTF model has been adopted and applied in both different technologies and also in HIT [7, 8]. TTF examines both of the two factors, the technology characteristics and the task characteristics to order understanding appropriate the requirements of the task to improve the performance of the user [6].

TTF includes two constructs: task characteristics and technology characteristics, which influence the utilization and task performance [6]. TTF proves that if technology used provides features that fit the demands, then satisfactory performance will be achieved [9]. The task characteristics and technical characteristics affect TTF, resulting in an impact on system usage and also on user performance.

The paper organised as follows. The next section describes the background details of the proposed research. The section after that presents research methodology followed by the data analysis. Section 4 describes mmodified proposed framework f.ollowing by the discussion and conclusion of the paper

## 2 Background

The study of CDSS acceptance contributes significantly to revealing many of the barriers and advantages in adopting the system and provides a significant opportunity for the success of the implementation of the CDSS. The investigation and discovery of the factors that influence the acceptance of CDSS by the end-user are crucial to its successful implementation [10]. Several previous studies have indicated the need to conduct high-quality studies to determine the factors that influence the acceptance of CDSS by physicians. In a study by Arts et al. [11] regarding the acceptance and obstacles concerning using CDSS in general practice, their results indicated a need to conduct more research on this issue to have a much better understanding of CDSS features required by GPs and to direct suppliers and designers to produce more effective systems based on the demands and requirements of the end-user.

Understanding the aspects which contribute to technology acceptance by physicians in the healthcare industry is significant for ensuring a simple application of new technologies [12]. IS acceptance's motivation is connected directly to the concept that systems are capable of completing their daily activities [1].
Acceptance of the CDSS is crucial in order to provide better health care services, since if the user does not accept the technology, the non-acceptance may affect negatively the health care and well-being of patients [1]

### 2.1 Theoretical conceptual framework

On the basis of the study and revision of different acceptance models, this research proposes to integrate TTF with UTAUT as figure 1 shows. This seems to be an appropriate conceptual framework to provide a contribution and effective model able to identify the determinants that affect CDSS as well as distinguishing the determinants that influence the new technology in the domain of HIS.

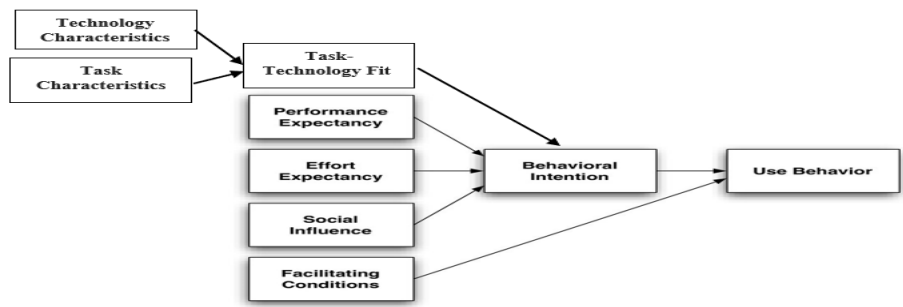

**Fig. 1.** Conceptual Model of Integration of UTAUT and TTF

Several studies have combined TTF with TAM [13, 14]. Usoro et al. [15] asserted that the combined both TAM and TTF together will help provide significant explanatory. In addition, several studies have combined TTF with UTAUT as well to investigate the technology acceptance [16, 17]

The integration of UTAUT and TTF frameworks will contribute considerably to identifying and discovering important factors which contribute to the understanding and investigation of user acceptance of technology. UTAUT and TTF have various advantages that help one to learn the factors that affect technology, so their combination contributes to achieving the most comprehensive advantages and benefits.

For understanding user acceptance in the technology of healthcare, we must comprehend not only the facts which affect acceptance but how these factors are fit as well. Even though various researches have explained the matter of 'fit', it is insufficient since its significance within the organization must be explored in detail, combining the technology with the user, to understand the issues which are concerned with the implementation of healthcare technology. There is actually a strong need to gain, address and understand the empirical support for the factor of fit when determining the acceptance of healthcare technologies by users [18] Researchers must examine the factors that affect user acceptance when it comes to the evaluation of the issue of user acceptance along with the factor the 'fit' among the technology and the users [18].

Khairat et al. [1] indicated that combining the models and frameworks would develop and enhance user acceptance to promote and assist the successful adoption of CDSS. They stated that if the user did not accept the system, there would be a lack of use of the technological system and, moreover, may threaten the healthcare and well-being of patients.

## 3    Methodology

This research employed a qualitative approach to collect data by conducting semi-structured interviews. Fifty-four interviews have been conducted with GPs in three stages to obtain their perspectives on and attitudes to the factors that influence the acceptance of CDSS. The procedure and implementation of three different stages in the qualitative approach contributes to increasing the level of validity and certainty of the data collected. The first stage initiates the factors' generation of the model, through convergent interviews where researchers interviewed 12 health professionals. The second stage: discovered and identified the factors of the model, by interviewing 42 GPs. These interviews helped the researchers to recognize the factors that influence the acceptance of CDSS through collecting perspectives, beliefs and attitudes of GPs towards CDSS. The third stage involved a review of the new proposed framework; researchers sought to increaser the validation of the final framework by discussing it with three GPs and the extent of their agreement and views about it.

Several studies had collected data based on the first and second stages in order to provide more accurate and detailed results for the phenomenon or issues studied [19, 20]. In this research, a third stage was added to have further investigation results of the proposed new framework.

### 3.1 Stage One: Initiated requirement generation of the framework

In Stage One, 12 exploratory interviews of GPs were conducted, using a convergent interviewing technique to gather insights and reasons for the factors behind using CDSS. In this stage, the UTAUT and TTF factors were reviewed and their appropriateness was reviewed to clarify and explore the appropriateness for the integration framework.

Convergent interviewing is a qualitative approach. It aims to collect, describe and understand individual preferences, attitudes and beliefs or to identify his or her significant interests [21]. The initial interviews in this approach help to make the questions more structured for the subsequent interviews. This enhances the credibility of the research. [22]. The convergent interview technique helped to recognise and identify the themes more easily and accurately [23]. This stage contributed to obtaining and discovering new factors by using convergent interviewing and devising questions based on previous interviews. The convergent technique was very relevant and valuable as it enabled the researcher to swiftly find the necessary issues and to establish the questions for the next stage [24].

### 3.2 Stage Two: Discovering the factors of the framework through perspectives

This stage provided significant data through 42 interviews with GPs. The questions in this stage related to the issues and factors mentioned and raised in the interviews of Stage One.

We used the case study approach to gather more data from 42 participants to explore and identify the factors that influence the acceptance of CDSS by GPs. This approach has been widely applied in several different fields of research, due to its distinctiveness and its contribution to obtaining valuable results [25, 26]. A case study collects data which greatly contributes to focusing on the research and identifies the issues [27]. Moreover, Gioia et al. [28] pointed out that such an approach provides opportunities for gaining insights into emerging concepts. This approach contributed to the exploration of new factors and the development of a proposal framework that explained the factors that influence the acceptance of CDSS by GPs. In-depth interviews led to the investigation of factors that influence the acceptance of CDSS. This also helps obtaining a broader understanding of perspectives and attitudes of the GPs toward adoption CDSS. The results of the in-depth interviews showed that all factors of both UTAUT and TTF influence the acceptance of CDSS by GPs, except social influence factor and the new discovered factors included Accessibility, Patient satisfaction, Communicability (With physicians), and Perceived Risk.

### 3.3 Stage Three: Validation of a new proposed framework

The third stage refers to reviewing and evaluating of the final framework with three physicians in order to obtain views and a final impression on the influencing factors that have been identified. This stage increased along with the second stage of validity the results and helps to gain a more comprehensive understanding of the final framework by the end-users of CDSS. The participants in this stage were among those 12 GPs who were interviewed in Stage One. This stage was added to obtain more views of the influencing factors from physicians because there new factors had been identified that had not been asked of them.

Herm et al. [29] indicated that reviewing the framework through interviews improves the validity of the framework. Maunder et al. [30] developed a framework for assessing the e-health readiness of dietitians. They conducted their study in three stages: a literature review, identification of topics related to the study, and interviews with 10 healthcare experts to verify and confirm the validity.

## 4    Data analysis

Thematic analysis was employed to analyse the data collected from the participants to understand and discover more about their experiences, perspectives, and attitudes regarding the factors that influence the acceptance of CDSS. The thematic analysis technique is widely applied in HIS studies [31]. Following this approach contributed to the formation of theories and models through a set of steps that assist to generate factors [32]. NVivo software has been used to analyse the data through applying six-step stages of thematic analysis established by Braun, Clarke [33]. This study followed those same phases to analyse the qualitative data which included: 1) Familiarising data; 2) Generating initial codes; 3) Searching for themes; 4) Reviewing themes; 5) Defining and naming themes, and 6) Producing the report. In Phase One (Familiarising data), the recording was reviewed more than once for analysing each interview's transcript to highlight the important issues and perspectives of the GPs. Phase Two (Generating initial codes), documents are coded according to their appropriate name in NVivo. Each code was linked into nodes to facilitate the process of building main and sub-themes. Phase Three (Searching for themes), After the initial arrangement and coding of the data, the symbols were classified into possible themes in addition to creating related sub-themes for the main themes. In Phase Four (Reviewing themes), the themes and their codes (established in the previous step) were checked and confirmed through comparing them with the interviews' transcripts. In Phase Five (Defining and naming themes), this step expresses the final access to the main themes, their identification and approval regarding their relevance to the codes. This prompted a comprehensive analysis of every theme and determined an illuminating or descriptive name for each theme. Phase Six: (Producing the report), a detailed explanation of each themes was undertaken to facilitate understanding each factor that influence the acceptance of CDSS.

## 5    Modified proposed framework

The study results showed that the following factors all influence the acceptance of CDSS by GPs. Performance Expectancy (Time, Alert, Accurate, Reduce Errors, Treatment Plan), Effort Expectancy and Facilitating Conditions (Training, Technical Support, Updating), Task-technology fit, Technology characteristics (Internet, Modern

Computers), Task characteristics, Accessibility, Patient satisfaction, Communicability (With physicians), and Perceived Risk (Time risk, Functional performance risk of the system)influence the acceptance of CDSS by GPs. These are shown in Figure 2.

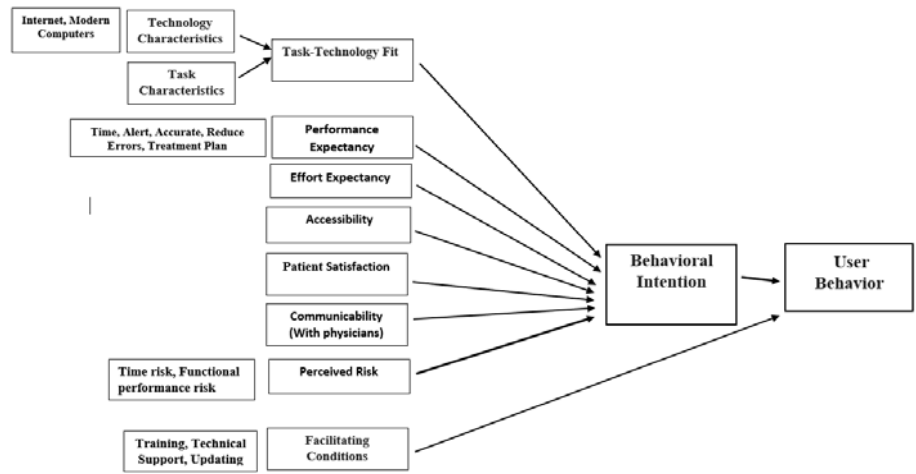

**Fig. 2.** Final modified framework.

The results contributed to gain an insight into the factors that influence acceptance and intention to use CDSS. Furthermore, more ideas and understanding of how to enhance the acceptance of CDSS and other advanced HIS systems were discovered and obtained.

## 6   Discussion and Conclusion

The CDSS is one of the advanced decision support mechanism that helps physicians to make more correct decisions using evidence-based facts or contents. Healthcare in developing countries needs more improve its practices in this aspect, in order to better understand the health protocols, to reduce medical errors and to provide better health care services. The framework developed in this research provides a new approach that helps to understand the factors that influence the use of CDSS. This will greatly benefit researchers, developers and providers of medical systems by way of designing more successful systems implementations. In addition, the new framework provides a better understanding regarding the features and tools in CDSS that help health professionals to provide quality and effective medical services and care.

Several HIS projects and systems have failed due to lack of consideration of the human side and the end-user considerations while designing health systems [34]. Analysis and determination of the requirements of the end-user of CDSS before its implementation and the final accreditation will save time, effort and money and will also contribute to the adoption of a successful HIS [35]. Furthermore, Kabukye et al. [36]

found that while health systems can improve health care, their adoption is still low because their systems do not meet the requirements of the user.

A limitation of this study is that this research relied on participants in Saudi Arabia as a developing country. The focus was mainly in two cities: Riyadh, which is the largest city in Saudi Arabia in terms of population and is also the capital, and Alqassim, as is it one of the closest areas to Riyadh (UN-Habitat, 2019). According to UN-Habitat, 2019, the population of Riyadh is 8,276,700 people, while the population of Qasim has 1,464,800 people .These cities were chosen due to travel and location restrictions in addition to the time and cost factors.

It was challenging to obtain enough participants due to the nature of their work and their concerns and being busy with patient care. The data collection process was interrupted due to longer waiting period for the GPs to agree to conduct the interview to obtain a suitable time for them. This research has only relied on a qualitative approach to explore the factors that affect GPs' acceptance of CDSS. Consequently, a quantitative approach was not suitable. The aim of this research is to build and develop theory instead of testing in a real healthcare decision domain [37]. Therefore, applying a qualitative approach through conducting semi-structured interviews is appropriate for this research. This research provides an opportunity for future research to study and verify the study's framework in studies that influence the acceptability of any new HIS design (for instance, using design science research [38, 39, 40]).

This research was conducted in Saudi Arabia through utilising the interview technique, so it may be possible to conduct other similar study using other research tools in other countries to determine if there are different or new factors. In addition, the focus of this study was on GPs, so other healthcare professionals would be of interests. It is possible to conduct further research that considers specialist or health professionals or consultants in different medical departments.